\documentclass[lettersize,journal]{IEEEtran}
\usepackage{amsmath,amsfonts,amsthm}
\usepackage{algorithmic}
\usepackage{algorithm}
\usepackage{array}
\usepackage[caption=false,font=normalsize,labelfont=sf,textfont=sf]{subfig}
\usepackage{textcomp}
\usepackage{stfloats}
\usepackage{url}
\usepackage{verbatim}
\usepackage{graphicx}
\usepackage{cite}
\usepackage{xcolor}

\DeclareMathOperator*{\argmax}{arg\,max}

\newtheorem{proposition}{Proposition}

\newtheorem{property}{Property}

\newcommand{\Co}{\overline{C}}

\let\oldnl\nl
\newcommand{\nonl}{\renewcommand{\nl}{\let\nl\oldnl}}

\begin{document}
	
	\title{\huge Smooth Actual EIRP Control for EMF Compliance\\ with Minimum Traffic Guarantees}
	
	\author{Lorenzo Maggi, Alo\"is Herzog, Azra Zejnilagic, Christophe Grangeat 
		
		\thanks{Lorenzo Maggi is with NVIDIA. Most of the work was carried out while he was at Nokia Bell Labs, 91377 Massy Palaiseau (France).  Alo\"is Herzog, Christophe Grangeat are with Nokia Networks France, Mobile Networks, 91377 Massy Palaiseau (France). Azra Zejnilagic is with Nokia Solutions and Networks, Mobile Networks, 89081 Ulm (Germany). Emails: lmaggi@nvidia.com, \{alois.herzog, azra.zejnilagic, christophe.grangeat\}@nokia.com}}
	
	
	
	\maketitle

	\begin{abstract}
		To mitigate Electromagnetic Fields (EMF) human exposure from base stations, international standards bodies define EMF emission requirements that can be translated into limits on the ``actual'' Equivalent Isotropic Radiated Power (EIRP), \textit{i.e.}, averaged over a sliding time window.
		We aim to enable base stations to adhere to these constraints while mitigating any impact on user performance. Specifically, our objectives are to: i) ensure EMF exposure compliance using actual EIRP control when implementing the ``actual maximum approach" described in IEC 62232:2022, ii) guarantee a minimum EIRP level, and iii) prevent resource shortages at all times.
		We first investigate exact and conservative algorithms, with linear and constant complexity, respectively, to compute the maximum allowed EIRP consumption under constraints i) and ii), referred to as EIRP ``budget''. Subsequently, we design a control method based on Drift-Plus-Penalty theory that preemptively curbs EIRP consumption only when needed to avoid future resource shortages.
		
	\end{abstract}

	
	
	\section{Introduction}
	\IEEEPARstart{T}{he} Radio Frequency (RF) Electromagnetic Field (EMF) limits for human exposure are established by international bodies in \cite{international2020guidelines}, \cite{ieee2019ieee}. 
	Considering the variability of EMF emissions from base stations, the ``actual maximum approach'' described in \cite{international2022determination} provides methods for assessing EMF compliance using time averaged, also called \emph{actual}, EMF emission values. These include the Equivalent Isotropic Radiated Power (EIRP), that can be controlled by the base station at runtime.
	According to channel modeling studies, setting a threshold on the actual EIRP at four times below the maximum EIRP achievable by the base station has a minimal impact on end-user performance \cite{baracca2018statistical}. However, lower thresholds may be applicable when employing higher directivity antennas \cite{rybakowski2023impact}.
	Among related works on EMF compliance, \cite{tornevik2020time} and \cite{wigren2021constrained} propose power control procedures limiting the usable bandwidth via model predictive control and optimal control theory, respectively; \cite{wigren2021coordinated} extends such approach to multiple radio transmitters. The work \cite{ying2015closed} optimizes precoding, while \cite{doose2017joint} addresses power control and precoding jointly. 
	The study in \cite{mandelli2024emf} discusses how to enforce the EIRP control at every slot, by fairly reducing the user transmitted power. 
	The Drift-Plus-Penalty (DPP) technique has been introduced in \cite{neely2010stochastic} and already applied to power control for energy savings in \cite{neely2015energy}, but not in the EMF domain.
	This letter investigates an actual EIRP control method via DPP guaranteeing a minimum EIRP level and preemptively reducing emissions to prevent resource shortages when implementing the actual maximum approach described in \cite{international2022determination}. Unlike the closest prior art \cite{tornevik2020time} that only reduces bandwidth, our method can be applied to any emission control method, including transmitted power, which has been shown in \cite{mandelli2024emf} to significantly reduce the impact on user performance.
	While presented for EIRP control, our method can be adapted to limit \emph{any} function of the emitted power. 

	\section{Problem formulation}

	The EIRP emitted by an antenna in the azimuth/elevation direction $(\phi, \theta)$ is the product of input power $P$ and antenna gain $G(\phi, \theta)$ with respect to an isotropic radiator:
	\begin{equation}
		\mathrm{EIRP} (\phi,\theta):=P \, G(\phi, \theta).
	\end{equation} 
	In free-space, the power density $S$ at distance $R$ from the radiation center of the antenna is proportional to the EIRP:
	\begin{equation}
		S(\phi, \theta) = \frac{\mathrm{EIRP}(\phi,\theta)}{4\pi R^2}  \ \ [\mathrm{W}/\mathrm{m}^2].
	\end{equation}
	
	We consider a discrete time grid indexed by $t$ and we denote the EIRP \emph{consumption} $c_t$ as the maximum EIRP over a predefined set of azimuth/elevation angles $\mathcal A$, called \emph{segment}: 
	\begin{equation} \label{eq:c_EIRP}
		c_t := \max_{(\phi, \theta)\in \mathcal A} \langle \mathrm{EIRP} (\phi,\theta)\rangle_t
	\end{equation}
	where $\langle.\rangle_t$ averages the EIRP over period $t$.
	Following \cite{international2022determination}, the \emph{actual} EIRP consumption, i.e., the average EIRP over a sliding window of $W$ periods, shall not exceed the configured threshold $\Co$:
	\begin{equation} \label{eq:EMFconstr}
		\frac{1}{W} \sum_{i=t-\min(t,W-1)}^{t} \!\!\! c_i(\gamma_i) \le \Co, \qquad \forall\, t\ge 0
	\end{equation}
	where $\gamma_t$ is our control variable, capping consumption $c_t$ as:
	\begin{equation} \label{eq:cap}
		0 \le c_t(\gamma_t) \le \gamma_t, \qquad \forall\, t\ge 0.
	\end{equation}
	To enforce \eqref{eq:cap} various methods can be used, such as reducing antenna transmission power \cite{mandelli2024emf}, transmission bandwidth \cite{tornevik2020time}, or beamforming gain \cite{ying2015closed}. Our solution, which involves controlling $\gamma$, is agnostic to the chosen power limitation method. 
	
	
	To guarantee a \emph{minimum service level}, we aim to keep the control $\gamma$ above the minimum EIRP level $\rho \Co$:
	\begin{equation} \label{eq:rhoC}
		\gamma_t \ge \rho \Co, \qquad \forall\, t\ge 0
	\end{equation}
	where $\rho\in [0,1]$ is the guaranteed EIRP ratio which is crucial in the presence of Guaranteed Bit-Rate (GBR) traffic. 
	
	Various control policies can jointly fulfill the constraints \eqref{eq:EMFconstr}-\eqref{eq:rhoC}. 
	The simplest one is a greedy policy that at each period $t$ sets $\gamma_t$ to the maximum allowed consumption $\Gamma_t$ under constraints \eqref{eq:EMFconstr}-\eqref{eq:rhoC}, later defined as EIRP \emph{budget}. 
	Such greedy policy is optimal in case of low traffic, where $c_t\ll \Gamma_t$. In fact, in this scenario, uncontrolled emissions are already below the configured threshold; therefore, any EIRP limitation would result in an unnecessary reduction in allocated resources, potentially leading to performance degradation. 
	Conversely, under the high traffic regime, setting $\gamma_t=\Gamma_t$ leads to budget depletion, after which the control drops to its minimum value $\rho \Co$. 
	One natural method to avoid such drops is to increase $\rho$. Yet, this leads to overly conservative policies: in fact, it turns out that when $\rho$ approaches 1, the budget $\Gamma_t$ tends to the actual EIRP threshold $\Co$.
	Thus, we study a method which decides whether and when preemptively curbing consumption to a value $\gamma_t<\Gamma_t$ to avoid any future budget depletion. 
	
	We formalize the desired preemptive control behavior as the one maximizing the average $\alpha$-fairness \cite{mo2000fair} of the control $\gamma$ over time, where the $\alpha$-fair function is defined as:
	\begin{equation}
		f(x):= \left\{ 
		\begin{array}{ll}
			\frac{x^{1-\alpha}}{1-\alpha} \quad \mathrm{if \ } \alpha\ge 0, \, \alpha\ne 1 \\
			\log(x) \quad \mathrm{if \ } \alpha=1
		\end{array} \right. .
	\end{equation}
	Hence, our problem becomes:
	\begin{align}
		\max_{\gamma} & \, \lim_{T\rightarrow \infty} \frac{1}{T} \sum_{t= 0}^{T-1} \mathbb E \left[ f(\gamma_t) \right] \label{eq:obj} \\
		\mathrm{s.t.} & \, \eqref{eq:EMFconstr},\eqref{eq:cap},\eqref{eq:rhoC} \notag 
	\end{align}
	where the expectation is with respect to the user demand. 
	Note that, if $\alpha=1$, $\alpha$-fairness coincides with proportional fairness; as $\alpha$ grows, it tends to max-min criterion \cite{mo2000fair}.
	
	Since $f(.)$ is concave, small values of $\gamma$ are severely penalized, thus encouraging $\gamma$ to evolve smoothly over time. 
	Moreover, since $f(.)$ is increasing, higher values of the control are preferred. Hence, the optimal control policy shall preemptively curb the EIRP consumption only ``when needed''.

	\section{EIRP budget}
	
	We start tackling our EIRP control problem \eqref{eq:obj} by characterizing its feasibility set, \textit{i.e.}, the set of control values $\gamma$ for which constraints \eqref{eq:EMFconstr}-\eqref{eq:rhoC} are fulfilled.
	
	\begin{proposition} \label{thm:feasible}
		Define the variable $\Omega_t$ as:
		\begin{equation} \label{eq:Omega}
			\Omega_t:=\max_{0\le k\le \min(t,W-1)} \sum_{i=1}^{k}\left(c_{t-i} - \rho \Co \right).
		\end{equation}
		To fulfill constraints \eqref{eq:EMFconstr}-\eqref{eq:rhoC}, the EIRP control $\gamma_t$ must satisfy:
		\begin{equation} \label{eq:feasible_set}
			\rho\Co\le \gamma_t \le \rho \Co + \Co(1-\rho) W - \Omega_t:=\Gamma_t, \quad \forall\, t\ge 0.
		\end{equation}
	\end{proposition}
	We denote the maximum feasible EIRP $\Gamma_t$ as \emph{EIRP budget}.

	\subsection{Budget computation}
	We now study how to compute the term $\Omega_t$, defining the budget $\Gamma_t$ via \eqref{eq:feasible_set}. 
	A naive approach would compute all $W$ terms in \eqref{eq:Omega} and take their maximum. Its complexity is \emph{quadratic} in the window length $W$. 
	Next we design more efficient methods.
	To this aim, we define the auxiliary variable:
	\begin{equation}\label{eq:Omega_tn}
		\Omega_t^n := \, \max_{k=0,\dots,n} \sum_{i=1}^k (c_{t-i}-\rho\Co), \quad 0\le n\le t, \ \forall\, t\ge 0
	\end{equation}
	and we call $\ell_t^n$ the argument of the maximum in \eqref{eq:Omega_tn}. 
	
	Observe that $\Omega_t:=\Omega_t^{\min(t,W-1)}$ and $\ell_t:=\ell_t^{\min(t,W-1)}$, where $\ell_t$ is the argument of maximum in \eqref{eq:Omega}. 
	
	We next prove a useful recursive property for $\Omega_t^n$.
	
	\begin{proposition} \label{prop:recursive_Omega}
		The following recursive relation holds:
		\begin{equation} \label{eq:omega_iter} 
			\Omega_{t+1}^{n+1} = \, \left[ \Omega_t^n + c_t - \rho\Co\right]^+, \quad n\le t, \ \forall\, t\ge 0.
		\end{equation}
	\end{proposition}
	
	Note that $[.]^+ := \max(.,0)$. It stems from Proposition \ref{prop:recursive_Omega} that, to compute $\Omega_t$, one can apply \eqref{eq:omega_iter} recursively over the last $W-1$ EIRP consumption samples, as described in Alg. \ref{alg:omega_t_scratch}. The resulting computational complexity is \emph{linear} in $W$.
	
	
	\begin{algorithm}[H]
		\caption{\emph{From-scratch} computation of \emph{exact} budget $\Gamma_t$} \label{alg:omega_t_scratch}
		
		Fix period $t$. Initialize $\Omega_t:=0$.\\
		\textbf{for} $i=\min(t,W-1),\dots,1$:
		\begin{itemize}
			\item [] Update $\Omega_t:= \left[ \Omega_t + c_{t-i} - \rho\Co \right]^+$
		\end{itemize}
		\textbf{return} $\Gamma_t = \rho \Co + \Co(1-\rho) W - \Omega_t$
		
	\end{algorithm}
	
	%

	\subsection{Iterative computation}
	
	Alg. \ref{alg:omega_t_scratch} computes $\Omega_t$ from scratch at each period $t$, \textit{i.e.}, it does not reuse computations performed at previous periods. 
	Yet, under some conditions, $\Omega_{t+1}$ can be computed from $\Omega_t$ with \emph{constant} (and negligible, in practice) complexity. This hinges on the following properties of variables $\Omega,\ell$.
	
	\begin{property} \label{prop:recurs_algo}
		If $\ell_t<W-1$, then $\Omega_{t+1}=\left[\Omega_t+c_t-\rho\Co\right]^+$.
	\end{property}

	\begin{property} \label{prop:recursive_ell}
		If $\Omega_{t+1}^{n+1}>0$, then $\ell_{t+1}^{n+1}=\ell_t^n + 1$. Else, $\ell_{t+1}^{n+1}=0$. 
	\end{property}

	\begin{property} \label{prop:TB_update}
		If $c_{t-i}\ge \rho \Co$ for $0\le i\le W-1$, then $\Omega_{t+1} = \Omega_t + c_t - c_{t-W+1}$.
	\end{property}
	
	Properties \ref{prop:recurs_algo}--\ref{prop:TB_update} suggest the following method for budget update. If the last $W$ consumption samples exceed $\rho \Co$, then $\Omega_{t+1}$ can be updated via Property \ref{prop:TB_update}. 
	Else, if $\ell_t<W-1$, then $\Omega_{t+1}$ can be derived as in Property \ref{prop:recurs_algo}. In both cases, the incremental complexity is \emph{constant} with respect to $W$. 
	Else, if $\ell_t=W-1$, then $\Omega_{t+1}$ is computed from scratch via the original Alg. \ref{alg:omega_t_scratch}. Finally, the value of $\ell_t$ is updated via Property \ref{prop:recursive_ell}. This method is formalized in Alg. \ref{alg:iterative}.

	\begin{algorithm}[H]
		\caption{\emph{Update} of \emph{exact} budget $\Gamma_t$}\label{alg:iterative}
		Initialize $\ell_0:=0,\, c_i:=0, \Omega_0:=0\ \forall\, i<0$.\\
		\textbf{for} period $t\ge 0$:
		\begin{itemize}
			\item[] \textbf{if} $c_{t-i}\ge \rho\Co, \ \forall \, i=0,\dots,W-1$ \textbf{then}: \begin{itemize}
				\item[] Set $\Omega_{t+1} = \Omega_t + c_t - c_{t-W+1}$
			\end{itemize}
			\item[] \textbf{else if} $\ell_t<W-1$ \textbf{then}:  $\Omega_{t+1}=\left[\Omega_t+c_t-\rho\Co\right]^+$
			
			\item[] \textbf{else}: Compute $\Omega_{t+1}$ via Alg. \ref{alg:omega_t_scratch}
			
			\item[] \textbf{if} $\Omega_{t+1}>0$ \textbf{then}: $\ell_{t+1}=\ell_t+1$
			
			\item[] \textbf{else}: $\ell_{t+1}=0$
			
			\item[] \textbf{return} $\Gamma_t = \rho \Co + \Co(1-\rho) W - \Omega_t$
		\end{itemize}
	\end{algorithm}

	\subsection{Low-complexity conservative approximation}
	
	The \emph{worst-case} complexity of iterative Alg. \ref{alg:iterative} is still \emph{linear} in $W$, although its \emph{average} complexity is lower than that of Alg. \ref{alg:omega_t_scratch}.
	The natural question is whether one can do better, especially if computational resources are a bottleneck.
	
	We here propose an alternative iterative method that provides a \emph{conservative} estimate of the EIRP budget with \emph{constant} complexity, \textit{i.e.}, not depending on the window length $W$.
	
	First, it is convenient to introduce the variable $\widetilde{\Omega}_t$, which is a variant of the original $\Omega_t$ defined in \eqref{eq:Omega} and is computed as the sum of the consumption excess with respect to $\rho\Co$ over the sliding window:
	\begin{equation}
		\widetilde{\Omega}_t := \sum_{i=1}^{\min(t,W-1)} \left[c_{t-1}-\rho\Co \right]^+.
	\end{equation}
	The associated feasibility set, analogous to \eqref{eq:feasible_set}, is defined as:
	\begin{equation} \label{eq:feasible_set_mod}
		\rho\Co\le \gamma_t \le \rho \Co + \Co(1-\rho) W - \widetilde{\Omega}_t:= \widetilde{\Gamma}_t, \quad \forall\, t\ge 0.
	\end{equation}
	Thus, $\widetilde{\Gamma}_t$ represents the maximum feasible control under the assumption that all past consumption values below the minimum guaranteed ratio $\rho\Co$ are \emph{overestimated} as $\rho\Co$. Consequently, it is expected that the approximated budget is generally \emph{lower} than the exact budget, as we formalize next.
	
	\begin{proposition} \label{prop:legacy}
		$\widetilde{\Gamma}_t\le \Gamma_t$, for all periods $t\ge 0$.
	\end{proposition}
	
	Thus, our approximate method produces \emph{overly conservative} policies, reducing EIRP consumption more than necessary. 
	
	On the positive side, the budget can be updated at each period via a simple rule with \emph{constant} complexity, requiring to add the latest consumption excess and subtracting the one exiting the new sliding window, as formalized in Alg. \ref{alg:legacy}.

	\begin{algorithm}[H]
		\caption{\emph{Update} of \emph{conservative} budget $\widetilde{\Gamma}_t$} \label{alg:legacy}
		Initialize $\widetilde{\Omega}_0:=0, \ c_i:=0, \ \forall\, i<0$.\\
		\textbf{for} period $t\ge 1$:
		\begin{itemize}
			\item[] Compute $\widetilde{\Omega}_{t} = \widetilde{\Omega}_{t-1} + \left[c_{t-1} -\rho\Co\right]^+ - \left[c_{t-W} -\rho\Co\right]^+$
			\item[] \textbf{return} $\widetilde{\Gamma}_t = \rho \Co + \Co(1-\rho) W - \widetilde{\Omega}_t$
		\end{itemize}
	\end{algorithm}

	\section{Smooth EIRP control}
	
	After characterizing the feasible EIRP control set, we now turn to the solution of the original problem \eqref{eq:obj}. 
	
	Two natural strategies arise. One approach is a greedy strategy, where $\gamma_t$ is simply set to $\Gamma_t$ (or $\gamma_t=\widetilde{\Gamma}_t$, if the conservative budget is used) for all $t$'s. Here, EIRP is constrained only when necessary. 
	However, under high traffic conditions, the budget can deplete, leading to $\gamma$ being pushed down to its minimum value $\rho\Co$ until the budget refills. 
	Alternatively, to prevent resource shortages, one can cautiously set $\gamma_t=\Co$ for all $t$'s. In this case, the control is consistently smooth, but low. 
	While optimal for high-load conditions, this cautious strategy proves overly conservative when traffic is low and EIRP control is unnecessary. 
	
	Ideally, the optimal strategy would not restrict EIRP during low traffic, while proactively reducing EIRP as traffic increases or when the budget is limited to prevent future resource shortages. 
	To effectively control $\gamma$, we propose a method based on Drift-Plus-Penalty (DPP) theory.

	\subsection{Preliminaries on Drift-Plus-Penalty (DPP)}
	
	Drift-Plus-Penalty (DPP) is a stochastic optimization technique typically employed to maximize a network utility while guaranteeing queue stability \cite{neely2010stochastic}. Formally speaking, DPP solves problems of the following form:
	\begin{align}
		\min_{x_t\in A(\omega_t)} & \, \lim_{T\rightarrow \infty} \frac{1}{T} \sum_{t=0}^T \mathbb E \left[ P(x_t,\omega_t) \right]:=p^* \label{eq:lyap} \\
		\mathrm{s.t.} & \, \lim_{T\rightarrow \infty} \frac{1}{T} \sum_{t=0}^{T-1} \mathbb E \left[ Y(x_t, \omega_t) \right] \le 0 \label{eq:lyap1}
	\end{align}
	where $x_t$ is the control variable, $\{\omega_t\}_t$ are \textit{i.i.d.} random variables, $P,Y$ are deterministic functions and $A(\omega_t)$ is the feasible control set. 
	By defining the \emph{virtual queue} $Q$ evolving as $Q_{t+1} = \left[ Q_t + Y(x_t, \omega_t) \right]^+$ for all $t\ge 0$, the constraint \eqref{eq:lyap1} can be equivalently formulated as $Q$ being mean-rate stable, \textit{i.e.}, $\lim_{t\rightarrow \infty} \mathbb E[Q_t]/t = 0$. 
	The DPP control is computed as:
	\begin{equation} \label{eq:alpha_dpp}
		x_t^{*} = \operatorname{argmin}_{x\in A(\omega_t)} Q_t Y(x,\omega_t) + V P(x,\omega_t)
	\end{equation} 
	which jointly reduces the queue Lyapunov drift (defined as the increment $Q_{t+1}^2-Q_t^2$) and the objective function. 
	The parameter $V> 0$ regulates the trade-off between the achieved performance, away from the optimal $p^*$ by $\mathcal O(V)$, and the average size $\mathcal O(1/V)$ of the virtual queue, as shown in \cite{neely2010stochastic}.

	\subsection{DPP-based heuristic}
	
	We can draw a \emph{partial} parallel between our problem \eqref{eq:obj} and the DPP formulation (\ref{eq:lyap}-\ref{eq:lyap1}) by setting the control $x_t:=\gamma_t$, the objective function $P(x_t,\omega_t):=-f(\gamma_t)$ and the function $Y(x_t,\omega_t):=c_t(\gamma_t)-\Co$.
	In this case, \eqref{eq:alpha_dpp} becomes:
	\begin{equation} \label{eq:gamma_dpp1}
		\min_{\gamma\in[\rho\Co,\Gamma_t]} \overline{Q}_t c_t(\gamma) - V f^{(\alpha)}(\gamma)
	\end{equation}
	where the virtual queue $\overline{Q}$ evolves as follows:
	\begin{equation} \label{eq:barQ}
		\overline{Q}_{t+1} = \left[ \overline{Q}_t + c_t(\gamma_t) - \Co \right]^+, \quad \forall\, t\ge 0
	\end{equation}
	with $\overline{Q}_0=0$. 
	Yet, \eqref{eq:gamma_dpp1} cannot be computed in practice, since $c_t(\gamma)$ is an \emph{unknown} function of $\gamma$. 
	Indeed, the variable $\omega_t$ (which can be interpreted as the consumption $c_t$ if \emph{no} power control is applied, \textit{i.e.}, $\gamma_t=\infty$) is unknown. 
	We will then bound \eqref{eq:gamma_dpp1} from above by replacing $c_t(\gamma)$ with $\gamma$ via \eqref{eq:cap} and minimize $\overline{Q}_t \gamma - V f^{(\alpha)}(\gamma)$ instead, leading to the expression:
	\begin{equation} \label{eq:gamma_dpp_expl}
		\gamma_t^* = \min \left( \max \left( V/\overline{Q}_t^{1/\alpha}, \, \rho\Co \right), \, \Gamma_t \right), \quad \forall \, t\ge 0.
	\end{equation}
	
\begin{figure*}
		\setlength\abovecaptionskip{-0.3\baselineskip}
		\centering
		\includegraphics[width=\linewidth]{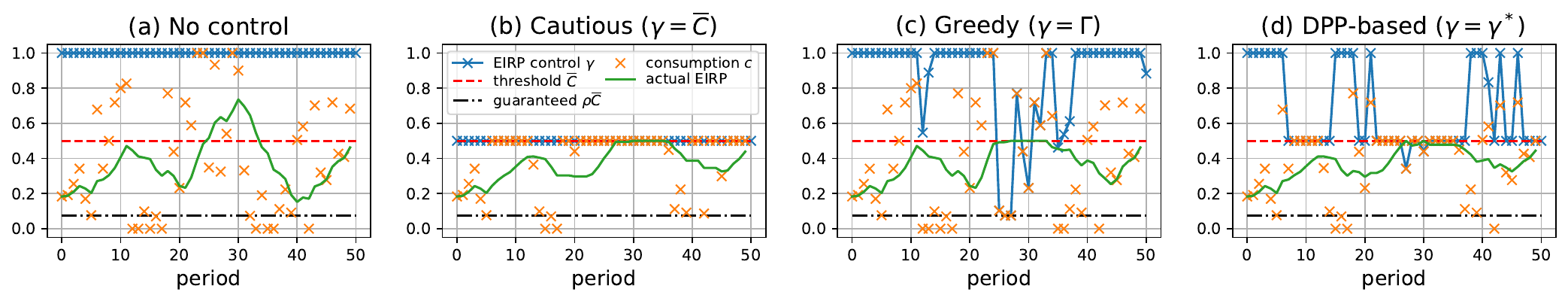}
		\caption{Illustrative example of the behaviour of multiple EIRP control strategies---(b) cautious, (c) greedy and (d) the proposed DPP---on EIRP consumption under identical input traffic, shown in (a). Parameters: $W=10,\rho=0.15,\alpha=1,\beta=0.95, V=15$.}
		\label{fig:comparison}
	\end{figure*}

	Note that, as traffic increases and EIRP consumption consistently surpasses the threshold $\Co$, the virtual queue grows, leading to a decrease in $\gamma$. By preemptively curbing consumption before the budget depletes, we prevent resource scarcity in the future, as illustrated in Fig. \ref{fig:comparison}.
	
	A second discrepancy between formulations \eqref{eq:obj} and \eqref{eq:lyap}-\eqref{eq:lyap1} lies in the averaging window, which spans $W$ samples in \eqref{eq:obj} and has an infinite duration in \eqref{eq:lyap}-\eqref{eq:lyap1}. It turns out that this leads the virtual queue $\overline{Q}$ to empty at most every $W$ samples in our finite window scenario, as stated below.
	\begin{proposition} \label{prop:Q0}
		For any list $\tau=[t',\dots,t+W-1]$ of $W$ successive periods, there exists a period $t\in \tau$ where $\overline{Q}_t=0$.
	\end{proposition}
	
	As the virtual queue empties, the control $\gamma^*_t$ as defined in \eqref{eq:gamma_dpp_expl} peaks at the budget $\Gamma_t$ at most every $W$ samples. 
	This can be particularly problematic in the case of high traffic, where any value $\gamma_t>\Co$ results in future budget depletion.
	To avoid this behavior, we redefine the virtual queue as follows:
	\begin{equation} \label{eq:barQ_mod}
		\overline{Q}^{\beta}_{t+1} = \left[ \overline{Q}^{\beta}_t + c_t(\gamma_t) - \beta \Co \right]^+, \quad 0\le \beta < 1, \  \forall\, t\ge 0
	\end{equation}
	where $\beta$ artificially inflates the virtual queue, thus preventing the undesirable behavior demonstrated in Proposition \ref{prop:Q0}.
	We describe our DPP-based EIRP control method in Alg. \ref{alg:DPP}.

	\begin{algorithm}[H]
		\caption{DPP-based power control for EMF compliance}\label{alg:DPP}
		Initialize $\overline{Q}^{\beta}_0:=0$.\\
		\textbf{for} period $t\ge 0$:
		\begin{itemize}
			\item[] Compute the budget $\Gamma_t$ via Alg. \ref{alg:iterative}
			\item[] \textbf{return} ${\gamma_t^{\beta}}^*=\min( \max ( V/(\overline{Q}^\beta_t)^{1/\alpha}, \, \rho\Co ), \, \Gamma_t )$ 
			\item[] Assess consumption $c_t$ and compute $\overline{Q}^{\beta}_{t+1}$ as in \eqref{eq:barQ_mod}
		\end{itemize}	
	\end{algorithm}

	\subsection{Numerical evaluations}
	
	For our experiments, we generated a series of traffic demands of $d_t$ bits. Demands are sparse: with probability $\ell$ (called \emph{load}), $d_t$ is drawn from a Zipf distribution, else $d_t=0$.
	The control $\gamma_t$ caps the requested EIRP at period $t$. The unserved demand is buffered and reappears in the next period.
	
	Fig. \ref{fig:comparison} shows the behavior of multiple EIRP control methods under the same traffic demand shown in (a). Cautious method (b) sets a constant control $\gamma=\Co$ and limits EIRP even under light load. Conversely, greedy (c) only controls EIRP once budget depletes, leading to resource shortage at periods $t\approx 25$. Our DPP-based method (d) limits the EIRP less frequently than "cautious" method, while still avoiding resource shortage under high load. DPP control may still drop temporarily below the threshold $\Co$, as seen at $t=28$: only an ideal oracle knowing future requests could consistently avoid this behavior.
	
\begin{figure}
		\setlength\abovecaptionskip{-0.3\baselineskip}
		\centering
		\includegraphics[width=\linewidth]{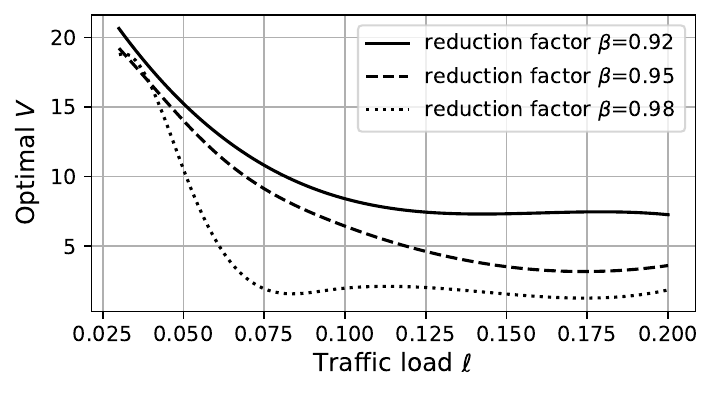}
		\caption{Value of parameter $V$ maximizing proportional fairness as the traffic load $\ell$ varies.}
		\label{fig:load_vs_Vopt}
	\end{figure}

	Fig. \ref{fig:load_vs_Vopt} shows how the optimal value for $V$, maximizing the proportional fairness ($\alpha=1$) criterion, varies with the traffic load $\ell$. Results are averaged over 100 independent traffic demand realizations. In the high load regime, DPP tends to behave as the "cautious" method since emissions must be preemptively curbed to avoid future budget depletion, leading to a decrease in the optimal $V$.
	Conversely, if the load is small ($\ell\approx 0$), it is optimal to curb emissions only when the budget depletes, resulting in a higher $V$. In this case, the DPP behavior is similar to the greedy one. 
	Since $\beta$ and $V$ have opposite effects on the control $\gamma$, decreasing $\beta$ makes the policy more conservative, while the optimal value of $V$ increases. 
	
	\begin{figure}
		\setlength\abovecaptionskip{-0.3\baselineskip}
		\centering
		\includegraphics[width=\linewidth]{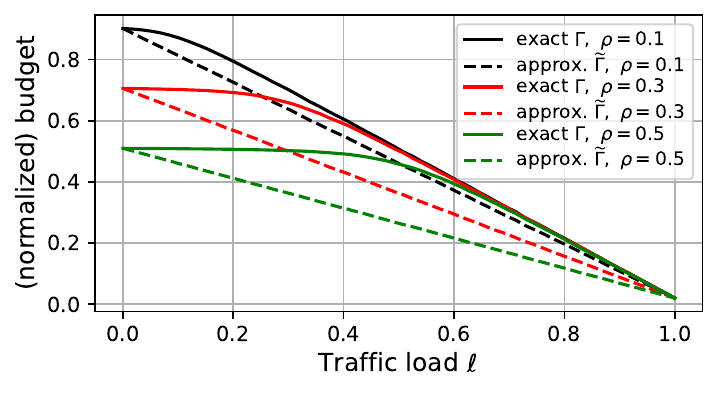}
		\caption{Exact budget $\Gamma$ versus conservative $\widetilde{\Gamma}$ as the traffic load $\ell$ varies.}
		\label{fig:budget_exact_vs_approx}
	\end{figure}

	Fig. \ref{fig:budget_exact_vs_approx} illustrates the relationship between the exact and the conservative budget, $\Gamma$ and $\widetilde{\Gamma}$, respectively. As load decreases ($\ell\approx 0$) or increases ($\ell>0.3$), the two solutions tend to coincide, as theory confirms. 
	Indeed, as consumption $c$ vanishes, $\Omega$ and $\widetilde{\Omega}$ both tend to zero; conversely, as consumption $c$ increases and is consistently higher than the guaranteed $\rho\Co$, the approximate excess $(c-\rho\Co)^+$ equals the actual excess $c-\rho\Co$, resulting in an exact budget, \textit{i.e.}, $\widetilde{\Gamma}=\Gamma$.
	However, in the middle load regime, the exact budget $\Gamma$ exceeds the conservative one $\widetilde{\Gamma}$. In fact, when consumption fluctuates around $\rho \Co$, only the exact solution is able to account for the past ``unused'' (positive) consumption deficit $\rho\Co - c$ to produce the current budget, while $\widetilde{\Gamma}$ approximates the deficit as null, leading to an overly conservative budget.

	\section{Conclusions}
	
	In this letter we propose an actual EIRP control method ensuring compliance with EMF exposure limits established in \cite{international2020guidelines}, \cite{ieee2019ieee} and using the actual maximum approach described in \cite{international2022determination}. We ensure that the \emph{actual} EIRP, averaged over a sliding time window, remains below a configured threshold. Additionally, our method consistently guarantees a minimum EIRP level, crucial for services like guaranteed bit-rate traffic, and proactively limits emissions to prevent future network resource shortage. 
	To compute the maximum allowed EIRP consumption, we devise an exact procedure with linear complexity (Alg. \ref{alg:iterative}) and a conservative one (Alg. \ref{alg:legacy}) with constant complexity. Our actual EIRP control algorithm, rooted in Drift-Plus-Penalty theory, preemptively limits the EIRP based on the status of a virtual queue, assessing the recent EIRP overshoot relative to the actual EIRP threshold $\Co$. The performance of our control method depends on parameters $V$ and $\beta$, which can be adjusted as the traffic load profile varies.

	\section*{Appendix: Proofs}
	
	\subsection{Proposition \ref{thm:feasible}}
	
	\begin{proof}
		The first inequality in \eqref{eq:feasible_set} stems from \eqref{eq:rhoC}. Next we fix period $t$ and assume $\gamma_t\le \Gamma_t$. We show that all the constraints \eqref{eq:EMFconstr} containing period $t$ can be satisfied by \emph{some} control policy. Set $\gamma_{t'}=\rho\Co$ for all $t'>t$. Then, since $c_t \le \gamma_t$, 
		$c_t \le \Gamma_t = \min_{0\le k\le W\!-\!1} W\Co -  \sum_{i=1}^{\min(t,k)} \!\!\! c_{t-i} - (W\!-\!k\!-\!1)\rho \Co$
		
		$\le \min_{0\le k\le W\!-\!1} W\Co - \sum_{i=1}^{\min(t,k)} c_{t-i} - \sum_{i=1}^{W-k-1} c_{t+i}$,\\
		which includes all constraints in \eqref{eq:EMFconstr} containing period $t$.
		Conversely, if $\gamma_t>\Gamma_t$, then by setting $\gamma_{t'}=\rho \Co$ for all $t'>t$, at least one constraint \eqref{eq:EMFconstr} is violated if $c_{t'}=\gamma_{t'}$ for all $t'>t$. Then, to satisfy \eqref{eq:EMFconstr} one must set $\gamma_{t'}<\rho \Co$ for some $t'>t$, which violates \eqref{eq:rhoC}. Thus, the upper bound is tight, \textit{q.e.d.}.
	\end{proof}

	\subsection{Proposition \ref{prop:recursive_Omega}}
	
	\begin{proof}
		We define $a_{t,k}:=\sum_{i=1}^k (c_{t-i}-\rho\Co)$. 
		By observing that $a_{t+1,k+1} = a_{t,k}+c_t-\rho\Co$, we derive that $\Omega_{t+1}^{n+1} = \max_{0\le k\le n+1} a_{t+1,k} = \max\left( a_{t+1,0}, \, \max_{0\le k\le n} a_{t+1,k+1} \right) = \big[ \max_{0\le k\le n} a_{t,k} + c_t - \rho\Co \big]^+=\big[\Omega_t^n + c_t - \rho\Co \big]^+$, \textit{q.e.d.}.
	\end{proof}

	\subsection{Property \ref{prop:recurs_algo}}
	\begin{proof}
		If $t<W-1$, then $\ell_t:=\ell_t^{\min(t,W-1)}=\ell_t^t<W-1$. It stems from Proposition \ref{prop:recursive_Omega} that $\Omega_{t+1}^{\min(t+1,W-1)} =  \Omega_{t+1}^{t+1} = \left[\Omega_t^t+c_t-\rho\Co\right]^+ 
		= \big[\Omega_t^{\min(t,W-1)}+c_t-\rho\Co \big]^+$.
		Next, let $t\ge W-1$. From Proposition \ref{prop:recursive_Omega} we know that  $\Omega_{t+1}^{W-1}=[\Omega_t^{W-2}+c_t-\rho\Co]^+$. If $\ell_t^{W-1}<W-1$, then 
		$\Omega_t^{W-1} :=  \max_{0\le k\le W-1} a_{t,k} 
		= \max_{0\le k\le W-2} a_{t,k}
		=: \Omega_t^{W-2}$, \emph{q.e.d.}.
	\end{proof}

	\subsection{Property \ref{prop:recursive_ell}}
	
	\begin{proof}
		It holds that $\ell_{t+1}^{n+1} = \argmax_{0\le k \le n+1} a_{t+1,k}$. 
		If $\Omega_{t+1}^{n+1}=0$, then $\ell_{t+1}^{n+1}=0$ since $a_{t+1,0}=0$. Else, if $\Omega_{t+1}^{n+1}>0$, then $	\ell_{t+1}^{n+1} = \argmax_{1\le k\le n+1} a_{t,k-1} + c_t-\rho\Co
		= \, \argmax_{1\le k\le n+1} a_{t,k-1}= \ell_t^n + 1$, \textit{q.e.d.}. 
	\end{proof}

	\subsection{Property \ref{prop:TB_update}}
	
	\begin{proof}
		If $c_{t-i}\ge \rho \Co$ for $0\le i\le W-1$, then it stems from \eqref{eq:Omega} that $\Omega_t=\sum_{i=1}^{W-1}(c_{t-i}-\rho\Co)$ and $\Omega_{t+1}=\sum_{i=0}^{W-2}(c_{t-i}-\rho\Co)$. The thesis stems from subtracting the last two expressions term by term.
	\end{proof}

	\subsection{Proposition \ref{prop:legacy}}
	
	\begin{proof}
		Since \eqref{eq:feasible_set_mod} holds, it suffices to prove that $\Omega_t\le \widetilde{\Omega}_t$. 
		Indeed, $\Omega_t \le \max_{0\le k\le \min(t,W-1)} \sum_{i=1}^{k} \left[ c_t - \rho\Co \right]^+ \!\!= \widetilde{\Omega}_t$.
	\end{proof}

	\subsection{Proposition \ref{prop:Q0}}
	
	\begin{proof}
		Let $t_1:=0< t_2<\dots$ be the periods at which $\overline{Q}=0$. Then, we will prove that $t_{i+1}-t_i\le W$ for all $i\ge 0$. Let us suppose that $\overline{Q}_{t_i}=0$. 
		We define $\Delta_i=\min\{ \delta: \, \sum_{j=0}^{\delta} (c_{t_i+j} - \Co) \le 0 \}$. It stems from \eqref{eq:EMFconstr} that $\Delta_i< W$. Also, $\overline{Q}_{t_i+\Delta_i} = \overline{Q}_{t_i}+\sum_{j=0}^{\Delta_i-1} (c_{t_i+j}-\Co)$. Thus, $\overline{Q}_{t_i+\Delta_i+1} = \left[ \overline{Q}_{t_i+\Delta_i} + (c_{t_i+\Delta_i} - \Co) \right]^+ = \big[ \overline{Q}_{t_i}+\sum_{j=0}^{\Delta_i} (c_{t_i+j}-\Co) \big]^+=0.$
		Therefore, $t_{i+1}-t_i=\Delta_i+1\le W$, \textit{q.e.d.}.
	\end{proof}


\end{document}